\documentclass[letterpaper, 10 pt, conference]{ieeeconf}  

\IEEEoverridecommandlockouts                              
\usepackage{amsmath} 
\usepackage{amssymb}  
\usepackage{graphicx,cite}

\usepackage[caption=false,font=footnotesize]{subfig}

\newcommand{\hs}{& \hspace{-2mm}}

\newcommand{\mc}[1]{\mathcal{#1}}
\newcommand{\Xa}{\mathcal{X}_{\rm a}}
\newcommand{\Prob}{\mathbb{P}}
\newcommand{\Exp}{\mathbb{E}}
\newcommand{\hP}{\hat{P}}
\newcommand{\hpi}{\hat{\pi}}
\newcommand{\hh}{\hat{h}}
\newcommand{\pih}{\pi^{\rm h}}
\newcommand{\Pih}{\Pi^{\rm h}}
\newcommand{\hPih}{\hat{\Pi}^{\rm h}}
\newcommand{\pim}{\pi^{\rm m}}
\newcommand{\Pim}{\Pi^{\rm m}}

\newcommand{\ul}[1]{\underline{#1}}

\usepackage{theorem}
\theorembodyfont{\rmfamily}

\newtheorem{prob}{Problem}
\newtheorem{lem}{Lemma}
\newtheorem{defin}{Definition}
\newtheorem{theorem}{Theorem}

\renewcommand{\QED}{\QEDopen}

\title{\LARGE \bf
Attack Impact Evaluation by Exact Convexification\\ through State Space Augmentation
}

\author{Hampei Sasahara, Takashi Tanaka, and Henrik Sandberg
\thanks{This work was supported by Swedish Research Council grant 2016-00861.}
\thanks{H. Sasahara is with the Department of Systems and Control Engineering, School of Engineering, Tokyo Institute of Technology, Tokyo 152-8552, Japan
        {\tt\small sasahara@sc.e.titech.ac.jp}}%
\thanks{T. Tanaka is with the Department of Aerospace Engineering and Engineering Mechanics, Cockrell School of Engineering, The University of Texas at Austin, TX 78712, USA
        {\tt\small ttanaka@utexas.edu}}%
\thanks{H. Sandberg is with the Division of Decision and Control Systems, School of Electrical Engineering and Computer Science, KTH Royal Institute of Technology, Stockholm SE-100 44, Sweden
        {\tt\small hsan@kth.se}}%
}

\begin{document}

\maketitle
\thispagestyle{empty}
\pagestyle{empty}

\begin{abstract}

We address the attack impact evaluation problem for control system security.
We formulate the problem as a Markov decision process with a temporally joint chance constraint that forces the adversary to avoid being detected throughout the considered time period.
Owing to the joint constraint, the optimal control policy depends not only on the current state but also on the entire
history, which leads to the explosion of the search space and makes the problem generally intractable.
It is shown that whether an alarm has been triggered or not, in addition to the current state is sufficient for specifying the optimal decision at each time step.
Augmentation of the information to the state space induces an equivalent convex optimization problem, which is tractable using standard solvers.
\end{abstract}

\section{INTRODUCTION}

Due to the increased connectivity, security of control systems has become an urgent matter.
Indeed, a lot of malicious software that target industrial control systems have been reported~\cite{History2018Hemsley}, and some of them succeeded to cause serious consequences to critical infrastructures~\cite{Nicolas2011Stuxnet,CISA2014,CISA2017,CISA2018}.
Security risk assessment is a necessary process for effective security countermeasures.
Risk assessment is conducted through specifying possible scenarios, quantifying their likelihoods, and evaluating their impacts~\cite{Kaplan1981On,Sridhar2012Cyber}.

Attack impact evaluation for control systems is a challenging task since it depends on malicious input sequences even if the supposed intrusion route is fixed.
Typically, the impact is quantified as the solution of an optimal control problem with a constraint that forces the adversary to avoid being detected throughout the considered time period~\cite{Teixeira2015Secure}.
Based on this framework, various formulations have been proposed~\cite{Mo2016On,Bai2017Data,Umsonst2017Security,Hirzallah2018Computation,Murguia2018Reachable,Chen2018Optimal,Teixeira2019Optimal,Milosevic2019Estimating,Fang2019Stealthy,Wang2020Optimal,
Sui2020The}.
A common problem is that the class of systems that can be handled by the existing works are limited because of difficulty to solve the constrained optimal control problem.
In particular, the class of attack detectors is limited.
Typical works consider the $\chi^2$ detector or provide a bound for all possible detectors by using Kullback-Leibler divergence (KLD) between the observed output and the nominal signal.
However, other stateful detectors such as cumulative sum (CUSUM) detectors and exponentially weighted moving average (EWMA) detectors are known to be effective in practice, from the both perspectives of detection performance and computational efficiency~\cite{Alvaro2011Attacks,Murguia2019On}.

The objective of this study is to provide an attack impact evaluation framework that can handle a general class of systems and detectors for practical risk assessment.
In our formulation, the control system with an attack detector is modeled as a Markov decision process (MDP) where an alarm region is embedded in the state space.
The stealth condition is given as a temporally joint chance constraint, which restricts the probability of intrusion into the alarm region throughout the whole time period.
However, because the chance constraint is joint over time, the optimal policy depends not only on the current state, but also the entire history.
In consequence, the dimension of the search space exponentially increases with the time horizon length, and the stochastic optimal control problem becomes generally intractable.

This paper proposes an equivalent problem formulation that reduces the size of the search space and makes the problem tractable.
It is shown that there exists a sufficient statistic for optimal decision making at each time step and we can reduce the size of the search space by augmenting the information into the state space.
Specifically, an extra binary state representing whether the alarm has been triggered or not in addition to the current state is sufficient for the optimal decision.
We refer to adding this information to the state space as \emph{state space augmentation}.
The optimal value can be achieved by Markov policies in the augmented MDP.
This finding induces an equivalent optimal control problem where the size of the search space is relatively small.
The induced problem is a standard constrained optimal control problem, which can be reduced to an equivalent linear program, and hence can be solved by standard solvers.
Moreover, it is observed that the problem formulation leads to an optimal policy where the adversary does not care about future alarms once an alarm is triggered.
In this case, the resulting state trajectory tends to stay in the alarm region, which is unreasonable in practice.
To model more sophisticated attack strategies, we propose an extended problem formulation by taking into account the number of alarms throughout the entire time period.
It is shown that state space augmentation can also be applied to the extended problem.

\subsection*{Related Work}

The attack impact evaluation problem has been widely considered in control system security~\cite{Teixeira2015Secure,Mo2016On,Bai2017Data,Umsonst2017Security,Hirzallah2018Computation,Murguia2018Reachable,Chen2018Optimal,Teixeira2019Optimal,Milosevic2019Estimating,Fang2019Stealthy,Wang2020Optimal,
Sui2020The}.
The formulations themselves are basically similar, and technically the problem is reduced to a constrained optimal control problem.
However, different approaches should be taken depending on the class of the system and the detector, the objective function that quantifies the impact caused by an attack signal, and the class of possible attack strategies.
An early work~\cite{Mo2016On} considers the $\chi^2$ detector, which is the simplest stateless detector.
The work~\cite{Bai2017Data} provides a rather general formulation where an essential bound is provided using KLD.
The CUSUM detector is considered in~\cite{Umsonst2017Security}.
In~\cite{Teixeira2019Optimal}, the author considers a detector using $\ell^2$ norm of the output signal assuming that the system is deterministic.
Finally, the other works consider one of the detectors.
To the best of our knowledge, this study is the first framework that can handle general detectors for attack impact evaluation.

Our proposed method is based on state space augmentation, which adds information of alarm history for decision making at each step.
A similar idea has been proposed in risk-averse MDP\cite{Bauerle2011Markov,Haskell2015Convex,Chow2017Risk}.
In~\cite{Bauerle2011Markov}, the authors consider a non-standard MDP where the objective function is given by not expectation but conditional-value-at-risk (CVaR), which is also referred to as average-value-at-risk.
They show that value iteration can be applied by considering the augmented state space even for CVaR.
The idea is generalized in~\cite{Haskell2015Convex} where not only CVaR minimization but also chance-constrained MDP are considered.
The work~\cite{Chow2017Risk} proposes risk-aware reinforcement learning based on the idea.
Moreover, linear temporal logic specification techniques provide a rather general framework~\cite{Baier2008Principles}.
While these works are relatively abstract, this study exhibits a clear interpretation through a concrete application in the context of security, where the augmented state has a concrete interpretation, namely the number of alarm's up until that time step.

\subsection*{Organization and Notation}
The paper is organized as follows.
Sec.~\ref{sec:prob} provides the system model, clarifies the threat model, and formulates the impact evaluation problem.
In Sec.~\ref{sec:conv}, the difficulty of the formulated problem is explained, and a solution based on state space augmentation is proposed.
Sec.~\ref{sec:ex} considers an extension of the problem where the number of alarms throughout the entire time period is taken into account.
It is shown that the approach proposed in Sec.~\ref{sec:conv} is still valid in the extended formulation.
In Sec.~\ref{sec:num}, the theoretical results are verified through numerical simulation.
Finally, Sec.~\ref{sec:conc} concludes and summarizes the paper.

Let $\mathbb{N}$ and $\mathbb{R}$ be the sets of natural numbers and real numbers, respectively.
The $k$-ary Cartesian power of the set $\mc{X}$ is denoted by $\mc{X}^k$.
The tuple $(x_0,\ldots,x_k)$ is denoted by $x_{0:k}$.

\section{ATTACK IMPACT EVALUATION PROBLEM}
\label{sec:prob}

\subsection{System Model}

This study considers a control system with an attack detector.
Its stochastic model is described by the finite-horizon MDP~\cite{Puterman1994Markov} with an alarm region given by
\begin{equation}\label{eq:MDP}
 \mc{M}:=(\mc{X},\mc{A},P,\mc{T},\{r_t\}_{t\in\mc{T}},\mc{X}_{\rm a})
\end{equation}
where
$\mc{X}$ is the state space,
$\mc{A}$ is the action space,
$P$ is the state transition function from $\mc{X}\times\mc{A}$ to $\mc{X}$,
$\mc{T}:=\{0,\ldots,T\}$ is the time index set,
$r_t:\mc{X}\times\mc{A}\to\mathbb{R}$ for $t=0,\ldots,T-1$ and $r_T:\mc{X}\to\mathbb{R}$ are the reward functions,
and $\mc{X}_{\rm a}\subset \mc{X}$ is the alarm region.
An alarm implemented in the control system is triggered when the state reaches the alarm region.
For simplicity, we assume the state space and the action space to be finite.

\begin{figure}[t]
  \centering
  \includegraphics[width=0.98\linewidth]{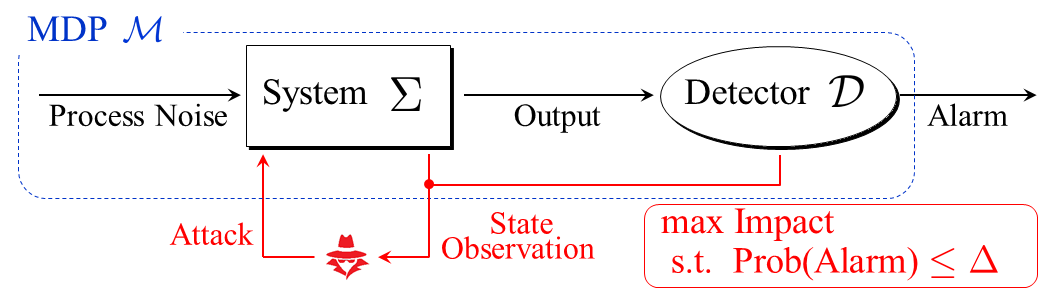}
  \caption{
  Block diagram of the stochastic system with a detector into which an attacker has intruded.
  The dynamics of the system with the detector is described as an MDP.
  The attacker aims at maximizing damage caused by the attack signal while avoiding triggering the alarm.
  }
  \label{fig:example}
\end{figure}

\emph{Example:}
We give a specific example that can be described in the form~\eqref{eq:MDP}.
Consider the block diagram in Fig.~\ref{fig:example}, which includes a nonlinear stochastic system equipped with a dynamic attack detector, which is also referred to as a stateful attack detector~\cite{Giraldo2018A}.
Let the dynamics of the control system $\Sigma$ and the detector $\mc{D}$ be given by
\begin{equation}\label{eq:ex_dyn}
 \Sigma: z_{t+1}=f(z_t,a_t,w_t),\quad
 \mc{D}: \left\{
 \begin{array}{ll}
 z^{\rm d}_{t+1} \hs = g(z^{\rm d}_t,z_t),\\
 \delta_t \hs = h(z^{\rm d}_t,z_t),
 \end{array}
 \right.
\end{equation}
respectively, where
$z_t\in\mc{Z}$ and $z^{\rm d}_t\in\mc{Z}^{\rm d}_t$ are states,
$a_t\in\mc{A}$ is an actuation signal caused by the attack,
$w_t\in\mc{W}$ is noise,
$\delta_t\in\{0,1\}$ is a binary signal that describes whether an alarm is triggered or not at the time step.
Suppose that the noise $w_t$ is an independent and identically distributed random variable over a probability space $(\Omega,\mc{F},\mathbb{P})$.
Let $\mc{X}:=\mc{Z}\times\mc{Z}^{\rm d}$ and $\mc{A}$ be the state space and the action space in the MDP form~\eqref{eq:MDP}, respectively.
Determine the state transition function by
\[
 \begin{array}{l}
  P((B_z,B^{\rm d}_z)|(z,z^{\rm d}),a)\\
  \quad := \left\{
  \begin{array}{ll}
  P_z(B_z|z,a) & {\rm if}\ g(z^{\rm d},z)\in B^{\rm d}_z,\\
  0 & {\rm otherwise},
  \end{array}
  \right.
 \end{array}
\]
where $B_z$ and $B^{\rm d}_z$ are Borel sets in $\mc{Z}$ and $\mc{Z}^{\rm d}$ and
\[
 \begin{array}{l}
  P_z(B_z|z,a):= \mathbb{P}(w_t\in B_w(B_z,z,a)), \\
  B_w(B_z,z,a):= \{w\in\mc{W}: f(z,a,w)\in B_z\}.
 \end{array}
\]
Further, let $\mc{X}_{\rm a}:=h^{-1}(\{1\})$ be the alarm region\footnote{Note that the state transition function $P$ is guaranteed to be a stochastic kernel if the maps $f,g,h$ are Borel measurable~\cite[Proposition 7.26]{Bertsekas1996Stochastic}.}.
Then, we can obtain an MDP in the form~\eqref{eq:MDP} by discretizing the spaces to be finite sets by using standard methods (for example, see~\cite{Munos2002Variable}).
More specific detector examples include the $\chi^2$ attack detector and the CUSUM attack detector~\cite{Murguia2019On}.
The $\chi^2$ attack detector with the observed output $y_t=Cz_t$, the nominal output $y^{\rm n}_t,$ and the threshold $\tau>0$ is represented by
\[
 \mc{D}: \delta_t=
 \left\{
 \begin{array}{ll}
 1 & {\rm if}\ \|Cz_t-y^{\rm n}_t\|^2 > \tau,\\
 0 & {\rm otherwise}.
 \end{array}
 \right.
\]
The CUSUM attack detector with the bias $b>0$ and the threshold $\tau>0$ is represented by
\[
 \mc{D}: \left\{
 \begin{array}{ll}
  z^{\rm d}_{t+1} \hs = \max (0,z^{\rm d}_t+\|Cz_t-y^{\rm n}_t\|-b),\\
  \delta_t \hs = \left\{
  \begin{array}{ll}
  1 & {\rm if}\ z^{\rm d}_t>\tau,\\
  0 & {\rm otherwise}
  \end{array}
  \right.
 \end{array}
 \right.
\]
with the observed signal and the nominal output\footnote{More precisely, the corresponding MDP should be time-varying owing to the nominal output. However, by adding the time information to the state, our discussion can readily be extended as long as finite-horizon problems are considered.}.

\subsection{Threat Model}
In this study, we consider the following threat model:

\begin{itemize}
\item The adversary has succeeded to intrude into the system and can execute any action $a\in\mc{A}$ in a probabilistic manner at every time step.
\item The adversary has perfect model knowledge of $\mc{M}$.
\item The adversary possesses infinite memory and computation resources.
\item The adversary can observe the state at every time step.
\item The attack begins at $t=0$ and ends at $t=T-1$.
\end{itemize}

The threat model implies that the adversary can implement an arbitrary history-dependent randomized policy $\pi\in\Pih$,
where $\pi=(\pi_t)_{t=0}^{T-1}$ is a tuple of policies at each time step,
$\Pih:=\{(\pi_t)_{t=0}^{T-1}: \pi_t:\mc{H}_t\to\mc{P}(\mc{A})\}$ is the set of all history-dependent randomized policies, $\mc{H}_t:=\mc{X}^t\times\mc{A}^{t-1}$ is the set of histories at the time step $t\in\mc{T}$,
and $\mc{P}(\mc{A})$ is the set of probabilistic measures with respect to the Borel algebra on $\mc{A}$.
Let the canonical measurable space of the MDP be denoted by $(\Omega,\mc{F})$, where $\Omega:=\mc{X}^{T+1}\times\mc{A}^T$ and $\mc{F}$ is its product $\sigma$-algebra.
Define the random variables $X_t$ and $A_t$ as the projection of each outcome into the state and action at the time step $t\in\mc{T}$, respectively.
Throughout this paper, the initial state is fixed for simplicity.
We denote the probability measure induced by $P$ given the policy $\pi$ by $\Prob^{\pi}$ and the expectation operator with respect to $\Prob^{\pi}$ by $\Exp^{\pi}$.

The objective of the adversary is to maximize a cumulative attack impact while avoiding being detected.
Let the impact be quantified as the expectation of the sum of the reward functions $r_t$ with respect to $t\in\mc{T}$.
Set the stealth condition to
\begin{equation}\label{eq:stealthy}
 \Prob^\pi(
 \lor_{t\in\mc{T}} X_t \in \Xa) \leq \Delta
\end{equation}
where
\[
 (\lor_{t\in\mc{T}} X_t \in \Xa):=\{x_{0:T} \in \mc{X}^{T+1}: \exists x_t\in\Xa \}
\]
with a constant $\Delta\geq 0$.
The criterion~\eqref{eq:stealthy} requires the probability of alarms to be less than or equal to $\Delta$ throughout the attack sequence.

\subsection{Problem Formulation}

Based on the setting above, the attack impact evaluation problem is formulated as a stochastic optimal control problem with a temporally joint chance constraint.
\begin{prob}\label{prob:ori}
The attack impact evaluation problem is given by
\begin{equation}\label{eq:prob_ori}
 \begin{array}{cl}
 \displaystyle{
 \max_{\pi\in\Pi^{\rm h}}} &
 \Exp^\pi
 \left[
 \sum_{t=0}^{T-1} r_t(X_t,A_t) + r_T(X_T)
 \right]
 \\
 {\rm s.t.} & \Prob^\pi(\lor_{t\in\mc{T}} X_t \in \Xa) \leq \Delta.
 \end{array}
\end{equation}
\end{prob}
Technically, the goal of this paper is to solve Problem~\ref{prob:ori}.
In the subsequent section, we explain its difficulty and propose a solution method.

\section{EXACT CONVEXIFICATION BY STATE SPACE AUGMENTATION}
\label{sec:conv}

\subsection{Idea: State Space Augmentation}
\label{subsec:idea}

It is well known that for standard MDPs without constraints the optimal value can be achieved by Markov policies, namely policies that depend only on the current state.
This property is justified by the fact: For any history-dependent policy, there exists a Markov policy such that the probabilities marginalized with respect to time are equal to those of the history-dependent policy~\cite[Theorem 18.1]{Hinderer1970Foundations}.
In Problem~\ref{prob:ori}, however, there exists a temporally joint chance constraint, which cannot be decomposed with respect to time.
Hence Markov policies cannot achieve the optimal value of~\eqref{eq:prob_ori} in general, an example of which is provided in Appendix~\ref{app:1}.
Thus the dimension of its search space exponentially increases with the time horizon length.
Indeed, the dimension of $\Pih$ is $\sum_{t\in\mc{T}}(|\mc{A}||\mc{X}|)^{t+1}$.
This explosion of the search space makes the problem intractable.

The key idea to overcome the difficulty is to add the alarm information by augmenting the state space.
We define the augmented state space and the induced augmented MDP.
\begin{defin}[Augmented State Space and MDP]
\label{def:aug}
The augmented state space of $\mc{X}$ for Problem~\ref{prob:ori} is defined as
\begin{equation}\label{eq:aug_SS}
 \hat{\mc{X}}:=\mc{X}\times\mc{Y}
\end{equation}
with
\[
 \mc{Y}:=\{0,1\}.
\]
The augmented MDP of $\mc{M}$ for Problem~\ref{prob:ori} is defined as
\begin{equation}\label{eq:aug_MDP}
 \hat{\mc{M}}:=(\hat{\mc{X}},\mc{A},\hat{P},\mc{T},\{\hat{r}_t\}_{t\in\mc{T}},\hat{\mc{X}}_{\rm a}),
\end{equation}
where the induced state transition function is given by
\[
 \begin{array}{l}
 \hP((x',0)|(x,0),a) = \left\{
 \begin{array}{ll}
 P(x'|x,a) & {\rm if}\ x'\not\in \Xa,\\
 0 & {\rm otherwise},
 \end{array}
 \right.\\
 \hP((x',0)|(x,1),a) = 0,\\
 \hP((x',1)|(x,0),a) = \left\{
 \begin{array}{ll}
 0 & {\rm if}\ x'\not\in \Xa,\\
 P(x'|x,a) & {\rm otherwise},
 \end{array}
 \right.\\
 \hP((x',1)|(x,1),a) = P(x'|x,a),
 \end{array}
\]
the induced reward function is given by $\hat{r}_t(x,y,a)=r_t(x,a)$ for any $y\in\mc{Y}$,
and the induced alarm region is given by $\hat{\mc{X}}_{\rm a}:=\Xa\times\mc{Y}$.
\end{defin}

The augmented state $y_t\in\mc{Y}$ represents the alarm information.
The binary flag $y_t=1$ indicates that the alarm has been triggered before the time step $t\in\mc{T}$, whereas $y_t=0$ indicates otherwise.
For the augmented MDP $\hat{\mc{M}}$, we denote the set of the history-dependent non-deterministic policies by $\hPih$,
the canonical measurable space by $(\hat{\Omega},\hat{\mc{F}})$,
the random variables of the state by $\hat{X}_t:=(X_t,Y_t)$,
the probability measure and the expectation induced by the initial condition $(x_0,0)$ and the policy $\hpi\in\hPih$ by $\Prob^{\hpi}$ and $\Exp^{\hpi}$, respectively.

By using the augmented information, we can rewrite the temporally joint chance constraint in~\eqref{eq:prob_ori} as an isolated chance constraint on the state at the final step.
It is intuitively true that $(\lor_{t\in\mc{T}} X_t \in \Xa)$, the event that an alarm is triggered at some time step, is equivalent to $Y_T^{-1}(\{1\})$, the event that the augmented state takes the value $1$ at the final time step.
This reformulation induces an equivalent problem
\begin{equation}\label{eq:prob_re}
 \begin{array}{cl}
 \displaystyle{
 \max_{\hpi\in\hat{\Pi}^{\rm h}}} &
 \Exp^{\hpi}
 \left[
 \sum_{t=0}^{T-1} r_t(X_t,A_t) + r_T(X_T)
 \right]
 \\
 {\rm s.t.} & \Prob^{\hpi}(Y_T=1) \leq \Delta.
 \end{array}
\end{equation}

The most important feature of this formulation is that the chance constraint depends on the marginal distribution only for the final time step.
Hence, the optimal value of~\eqref{eq:prob_re} can be achieved by Markov policies for the augmented state space $\hat{\mc{X}}$.
Thus, the problem~\eqref{eq:prob_re} can be reduced to
\begin{equation}\label{eq:prob_re2}
 \begin{array}{cl}
 \displaystyle{
 \max_{\hpi\in\hat{\Pi}^{\rm m}}} &
 \Exp^{\hpi}
 \left[
 \sum_{t=0}^{T-1} r_t(X_t,A_t) + r_T(X_T)
 \right]
 \\
 {\rm s.t.} & \Prob^{\hpi}(Y_T=1) \leq \Delta,
 \end{array}
\end{equation}
where the search space of~\eqref{eq:prob_re} is replaced with $\hat{\Pi}^{\rm m}$, the set of Markov policies for the augmented MDP $\hat{\mc{M}}$.

We formally justify the reformulation.
We first show the following lemma.
\begin{lem}\label{lem:joint}
For any $\hpi\in\hat{\Pi}^{\rm h}$, there exists $\pi\in\Pih$ such that
\begin{equation}\label{eq:joint_eq}
P^{\pi}_{X_{0:t},A_{0:t}}=P^{\hpi}_{X_{0:t},A_{0:t}},\quad \forall t\in\mc{T}
\end{equation}
where $P^{\pi}_{X_{0:t},A_{0:t}}$ is the joint distribution of $(X_{0:t},A_{0:t})$ for $\mc{M}$ with respect to $\Prob^{\pi}$ and $P^{\hpi}_{X_{0:t},A_{0:t}}$ is that for $\hat{\mc{M}}$ with respect to $\Prob^{\hpi}$.
\end{lem}
\begin{proof}
Let $\hh_t=(x_{0:t},y_{0:t},a_{0:t-1})$ be a history associated with the augmented MDP $\hat{\mc{M}}$.
Denote the first time instance at which $y_t=1$ by $\ul{t}$, i.e.,
$\ul{t}:= \min\{t \in \mc{T}: y_t=1\}$
and $\ul{t}:=\infty$ when $y_t=0$ for any $t\in\mc{T}$.
We say $\hh_t$ to be consistent when $\hh_t$ satisfies
\[
\left\{
 \begin{array}{ll}
 y_t=1, & \forall t\geq \ul{t},\\
 x_t \not \in \Xa, & \forall t < \ul{t},\\
 x_{\ul{t}} \in \Xa. & 
 \end{array}
\right.
\]
Note that the probability of any history that is not consistent is zero from the induced state transition function $\hat{P}$ for any policy.
A history $h_t$ in the original MDP $\mc{M}$ uniquely induces a consistent history in $\hat{\mc{M}}$, which we denote by $\hh_t(h_t)$.
For a fixed policy $\hpi_t$ for $\hat{\mc{M}}$, give a policy $\pi_t$ for $\mc{M}$ by
\begin{equation}\label{eq:eq_pol}
 \pi_t(a|h_t):=\hpi_t(a|\hh_t(h_t)),\quad t\in\mc{T}.
\end{equation}

We confirm that the policy~\eqref{eq:eq_pol} satisfies the condition~\eqref{eq:joint_eq} by induction.
Because the initial state is fixed to be $(x_0,0)$, the condition is satisfied for $t=0$.
Assume that~\eqref{eq:joint_eq} holds for some $t\in\mc{T}$.
Denoting the probability mass function with a policy $\pi$ by $P^{\pi}$, we have
\[
\begin{array}{l}
 P^{\hpi}(x_{0:t+1},a_{0:t+1}) \\
  = P^{\hpi}(x_{t+1},a_{t+1}|x_{0:t},a_{0:t})P^{\hpi}(x_{0:t},a_{0:t})\\
  = P^{\hpi}(a_{t+1}|x_{0:t+1},a_{0:t}) P(x_{t+1}|x_t,a_t) P^{\pi}(x_{0:t},a_{0:t})
\end{array}
\]
for any $x_{0:t+1}$ and $a_{0:t+1}$.
Thus, it suffices to show that
\begin{equation}\label{eq:eq_mass}
 P^{\hpi}(a_{t+1}|x_{0:t+1},a_{0:t}) = P^{\pi}(a_{t+1}|x_{0:t+1},a_{0:t}).
\end{equation}
For a policy $\hpi$, we have
\[
 \begin{array}{l}
 P^{\hpi}(a_{t+1}|x_{0:t+1},a_{0:t}) \\
 = \sum_{y_{0:t+1}\in\mc{Y}^{t+2}} \hpi(a_{t+1}|\hh_{t+1}) P^{\hpi}(y_{0:t+1}|x_{0:t+1},a_{0:t}).
 \end{array}
\]
Since $(x_{0:t+1},a_{0:t})$ uniquely induces $y_{0:t+1}$ such that the corresponding history becomes consistent and satisfies $P^{\hpi}(y_{0:t+1}|x_{0:t+1},a_{0:t})=1$,
we have
\[
 \begin{array}{cl}
 P^{\hpi}(a_{t+1}|x_{0:t+1},a_{0:t}) & = \hpi_{t+1}(a_{t+1}|\hh_{t+1}(h_{t+1}))\\
  & = \pi_{t+1}(a_{t+1}|x_{0:t+1},a_{0:t})
 \end{array}
\]
and~\eqref{eq:eq_mass} holds.
\end{proof}

Lemma~\ref{lem:joint} implies that the stochastic behaviors of the original MDP and the augmented MDP are identical with appropriate policies.

The following theorem is the main result of this paper.
\begin{theorem}~\label{thm:main}
Denote the optimal values of the chance-constrained optimal control problems in~\eqref{eq:prob_ori},~\eqref{eq:prob_re}, and~\eqref{eq:prob_re2} by $J^\ast,$ $\hat{J}^{\ast},$ and $\hat{J}^{{\rm m}\ast},$ respectively.
Then we have $J^\ast=\hat{J}^{\ast}=\hat{J}^{{\rm m}\ast}.$
\end{theorem}
\begin{proof}
We first show $J^\ast=\hat{J}^{\ast}$.
Since the policy set of the augmented MDP includes that of the original MDP, $J^\ast\leq\hat{J}^{\ast}$ clearly holds.
Fix a feasible policy $\hpi\in\hPih$ for~\eqref{eq:prob_re} and take the corresponding policy $\pi\in\Pih$ for the original MDP according to~\eqref{eq:eq_pol}.
From Lemma~\ref{lem:joint}, the marginal distributions of the state and the action with the policies coincide.
By denoting $(\lor_{t\in\mc{T}} X_t \in \Xa)$ by $\mc{E}^1_{\rm a}$, from~\eqref{eq:eq_pol}, we have $\Prob^{\pi}(Y_T=1|\mc{E}^1_{\rm a})=\Prob^{\pi}(\mc{E}^1_{\rm a}|Y_T=1)$.
Thus $\Prob^{\pi}(Y_T=1)=\Prob^{\pi}(\mc{E}^1_{\rm a})=\Prob^{\hpi}(\mc{E}^1_{\rm a})\leq \Delta$ and $\pi$ is feasible in~\eqref{eq:prob_re}.
Therefore $J^\ast\geq\hat{J}^{\ast}$, which leads to $J^\ast=\hat{J}^{\ast}$.
Finally, $\hat{J}^{\ast}=\hat{J}^{{\rm m}\ast}$ is a direct conclusion of~\cite[Theorem 18.1]{Hinderer1970Foundations}.
\end{proof}

Theorem~\ref{thm:main} justifies the reformulation from~\eqref{eq:prob_ori} to~\eqref{eq:prob_re2}.
The dimension of $\hat{\Pi}^{\rm m}$ is
\[
 |\mc{A}||\hat{\mc{X}}||\mc{T}|=|\mc{A}||\mc{X}||\mc{Y}||\mc{T}|=2|\mc{A}||\mc{X}||\mc{T}|,
\]
which is much smaller than $\sum_{t\in\mc{T}}(|\mc{A}||\mc{X}|)^{t+1}$, that of the history-dependent search space.
Finally, note that one only has to memorize the binary information $y_t$ for implementation of the optimal policy.


\subsection{Solution to Problem~\ref{prob:ori}}

It is known that a standard class of constrained MDPs, which the problem~\eqref{eq:prob_re2} belongs to, can be solved using linear programming~\cite{Altman1999Constrained}.
The procedure is briefly explained in the following.
Denote $\Prob^{\hpi}(\hat{X}_t=\hat{x},A_t=a)$ by $\rho_t(\hat{x},a)$ where $\hpi$ is omitted from the notation.
The objective function can be rewritten by
\[
 \begin{array}{l}
 \Exp^{\hpi}
 \left[
 \sum_{t=0}^{T-1} r_t(X_t,A_t) + r_T(X_T)
 \right]\\
 =
 \sum_{t=0}^{T-1} \sum_{\hat{x}\in\hat{\mc{X}},a\in\mc{A}} \hat{r}_t(\hat{x},a) \rho_t(\hat{x},a)
 + \sum_{\hat{x}\in\hat{\mc{X}}}\hat{r}_T(\hat{x})\rho_T(\hat{x}).
 \end{array}
\]
The left-hand side of the constraint can be rewritten by
\[
 \Prob^{\hpi}(Y_T=1)
 =\Exp^{\hpi}\left[
 \delta_{y=1}(\hat{X}_T)
 \right]
 =\sum_{\hat{x}\in\hat{\mc{X}}}\delta_{y=1}(\hat{x})\rho_T(\hat{x})
\]
where $\delta_{y=1}(\hat{x})$ for $\hat{x}=(x,y)$ takes $1$ if $y=1$ and $0$ otherwise.
Moreover, the function $\rho_t(\hat{x},a)$ satisfy
\begin{equation}\label{eq:meas_cond}
\left\{
\begin{array}{l}
\displaystyle{
 \sum_{a\in\mc{A}}\rho_t(\hat{x},a) =  \hspace{-4mm} \sum_{\bar{x}\in\hat{\mc{X}},a\in\mc{A}} \hspace{-2mm} \hat{P}(\hat{x}|\bar{x},a)\rho_{t-1}(\bar{x},a),\ 1\leq t\leq T-1,}\\
 \rho_0(\hat{x},a)=\delta_{x=x_0}(\hat{x}),\\
 \rho_T(\hat{x})= \sum_{\bar{x}\in\hat{\mc{X}},a\in\mc{A}} \hat{P}(\hat{x}|\bar{x},a) \rho_{T-1}(\bar{x},a),\\
 0\leq \rho_t(\hat{x},a)\leq 1,\quad \forall \hat{x}\in\hat{\mc{X}},\forall a\in\mc{A},t=0,\ldots,T-1,\\
 0\leq \rho_t(\hat{x})\leq 1,\quad \forall \hat{x}\in\hat{\mc{X}}
\end{array}\right.
\end{equation}
where $\delta_{x=x_0}(\hat{x})$ for $\hat{x}=(x,y)$ takes $1$ if $x=x_0$ and $0$ otherwise.
Conversely, for any functions that satisfy~\eqref{eq:meas_cond}, there exists a policy $\hpi\in\hat{\Pi}^{\rm m}$ such that $\Prob^{\hpi}(\hat{X}_t=x,A_t=a)=\rho_t(\hat{x},a)$.
Such a policy can specifically be constructed by
\begin{equation}\label{eq:pol_prob_trans}
\textstyle{
 \hpi^{\rm m}_t(a|\hat{x})=\rho_t(\hat{x},a)/\sum_{\bar{a}\in\mc{A}}\rho_t(\hat{x},\bar{a})
}
\end{equation}
when the denominator is nonzero, otherwise  $\hpi^{\rm m}_t(a|\hat{x})$ can be arbitrary.
Therefore $\rho_t(\hat{x},a)$ can be used as the decision variables of the problem~\eqref{eq:prob_re2} instead of the policy.
It should be emphasized that all the functions are linear with respect to $\rho_t(\hat{x},a)$.

The linear programming formulation is given as follows:
\begin{equation}\label{eq:LP}
\begin{array}{cl}
\displaystyle{
\max_{\rho}
}
&
\displaystyle{
\sum_{t=0}^{T-1} \sum_{\hat{x}\in\hat{\mc{X}},a\in\mc{A}} \hat{r}_t(\hat{x},a) \rho_t(\hat{x},a)
 + \sum_{\hat{x}\in\hat{\mc{X}}}\hat{r}_T(\hat{x})\rho_T(\hat{x})
} \\
 {\rm s.t.} &
\displaystyle{ 
 \sum_{\hat{x}\in\hat{\mc{X}}}\delta_{y=1}(\hat{x})\rho_T(\hat{x})\leq \Delta
 }\ {\rm and}\ \eqref{eq:meas_cond},
\end{array}
\end{equation}
which can readily be solved by standard solvers.

The solution to Problem~\ref{prob:ori} is summarized as follows:
\begin{enumerate}
\item Given the optimization problem~\eqref{eq:prob_ori}, consider the state space augmentation given by Definition~\ref{def:aug}.
\item Using the augmented MDP, formulate the linear programming in~\eqref{eq:LP}.
\item The optimal value of~\eqref{eq:LP} is equal to that of~\eqref{eq:prob_ori}. The optimal policy is obtained through~\eqref{eq:eq_pol} and~\eqref{eq:pol_prob_trans}.
\end{enumerate}

\section{EXTENSION: MULTI-ALARM AVOIDANCE STRATEGY}
\label{sec:ex}

The optimal policy of Problem~\ref{prob:ori} means that the adversary cares about being detected when there has been no alarms so far, but does no longer care once an alarm is triggered.
In practice, however, a single alarm may not result in counteractions by the defender owing to existence of false alarms.
Hence, it is more natural that the adversary avoids serial alarms.
In this sense, Problem~\ref{prob:ori} leads to an unreasonable strategy, which is illustrated by Fig.~\ref{fig:res1} in the simulation section.

The cause of such unnatural strategies is that the chance constraint is given for the binary value whether an alarm has been triggered or not in the problem formulation.
In this section, to obtain a more reasonable strategy and evaluate the attack impact more precisely, we extend the formulation of Problem~\ref{prob:ori} by taking multi-alarms into account.

First, define the event that alarms are triggered more than or equal to $i$ times by
\[
 \mc{E}^i_{\rm a}:=\{x_{0:T}\in\mc{X}^{T+1}: |\mc{T}_{\rm a}(x_{0:T})|\geq i \}
\]
where
\[
 \mc{T}_{\rm a}(x_{0:T}):=\{t\in\{0,\ldots,T\}:x_t\in\mc{X}_{\rm a}\}.
\]

Using the notation, the extended version of the attack impact evaluation problem for multi-alarm avoidance strategies is formulated as follows:

\begin{prob}\label{prob:mul}
The attack impact evaluation problem for multi-alarm avoidance strategies is given by
\begin{equation}\label{eq:prob_mul}
 \begin{array}{cl}
 \displaystyle{
 \max_{\pi\in\Pi^{\rm h}}} &
 \Exp^\pi
 \left[
 \sum_{t=0}^{T-1} r_t(X_t,A_t) + r_T(X_T)
 \right]
 \\
 {\rm s.t.} & \Prob^\pi(\mc{E}_{\rm a}^1) \leq \Delta_1\\
  & \quad \vdots\\
  & \Prob^\pi(\mc{E}_{\rm a}^T) \leq \Delta_T
 \end{array}
\end{equation}
where $\Delta_i\geq 0$ for $i=1,\ldots,T$ are constants for the stealth constraints.
\end{prob}

The same idea of state space augmentation can be applied to Problem~\ref{prob:mul} as well by adding information of the number of alarms instead of the binary information.
The augmented state space and the induced augmented MDP for Problem~\ref{prob:mul} are defined as follows.
\begin{defin}
The augmented state space of $\mc{X}$ for Problem~\ref{prob:mul} is defined as~\eqref{eq:aug_SS} with
\[
 \mc{Y} := \{0,\ldots,T\}.
\]
The augmented MDP for Problem~\ref{prob:mul} is defined as~\eqref{eq:aug_MDP} where the induced state transition function is given by
\[
 \begin{array}{l}
 \hP((x',y)|(x,y),a) = \left\{
 \begin{array}{ll}
 P(x'|x,a) & {\rm if}\ x'\not\in \Xa,\\
 0 & {\rm otherwise},
 \end{array}
 \right.\\
 \hP((x',y')|(x,y),a) = 0,\ \forall y'\not\in \{y, y+1\},\\
 \hP((x',y+1)|(x,y),a) = \left\{
 \begin{array}{ll}
 0 & {\rm if}\ x'\not\in \Xa,\\
 P(x'|x,a) & {\rm otherwise},
 \end{array}
 \right.
 \end{array}
\]
the induced reward function and the induced alarm region are given by those in Definition~\ref{def:aug}.
\end{defin}

The augmented MDP naturally leads to an equivalent problem
\begin{equation}\label{eq:prob_mul2}
 \begin{array}{cl}
 \displaystyle{
 \max_{\hpi\in\hat{\Pi}^{\rm m}}} &
 \Exp^{\hpi}
 \left[
 \sum_{t=0}^{T-1} r_t(X_t,A_t) + r_T(X_T)
 \right]
 \\
 {\rm s.t.} & \Prob^{\hpi}(Y_T\geq 1) \leq \Delta_1\\
  & \quad \vdots\\
  & \Prob^{\hpi}(Y_T\geq T) \leq \Delta_T
 \end{array}
\end{equation}
where the search space is the set of Markov policies.
The following theorem is the correspondence of Theorem~\ref{thm:main}.
\begin{theorem}
Denote the optimal values of the problems in~\eqref{eq:prob_mul} and~\eqref{eq:prob_mul2} by $J^{\ast}$ and $\hat{J}^{{\rm m}\ast}$, respectively.
Then we have
$J^{\ast}=\hat{J}^{{\rm m}\ast}.$
\end{theorem}
\begin{proof}
The claim can be proven in a manner similar to the proof of Theorem~\ref{thm:main}.
\end{proof}

The dimension of $\hat{\Pi}^{\rm m}$ is $|\mc{A}||\mc{X}||\mc{T}|^2$.
The problem~\eqref{eq:prob_mul2} is also a standard constrained MDP and can be reformulated as a linear program.

\emph{Remark:}
The constraint in the extended problem~\eqref{eq:prob_mul} can be regarded as a restriction on the probability distribution of the number of alarms.
In other words, the formulation utilizes a risk measure on a probability distribution.
Several risk measures have been proposed, such as CVaR, which is one of the most commonly used coherent risk measures~\cite{Artzner1999Coherent}.
Those risk measures compress risk of a random variable that possesses a distribution into a scalar value.
Because our formulation uses the full information of the distribution, the constraint can be regarded as a fine-grained version of constraints using standard risk measures.

\section{NUMERICAL EXAMPLE}
\label{sec:num}

We verify the effectiveness of the proposed method through a numerical example.
Consider the state space $\mc{X}=\{1,\ldots,16\}$.
The state $x_t=1$ represents the reference value, and the adversary's objective is to push the state from the reference as much as possible.
The action space is given by $\mc{A}:=\{{\rm up},{\rm stay}, {\rm down}\}$.
The time horizon is given by $\mc{T}=\{0,1,\ldots,15\}$.
The state transition function is given by
\[
 \begin{array}{l}
  \left\{
  \begin{array}{ll}
  P(x+1|x,{\rm up})\hs=0.8,\\
  P(x|x,{\rm up})\hs=0.2,
  \end{array}
  \right.
  \left\{
  \begin{array}{ll}
  P(x+1|x,{\rm stay})\hs=0.2,\\
  P(x|x,{\rm stay})\hs=0.8,
  \end{array}
  \right.\\
  \left\{
  \begin{array}{ll}
  P(x+1|x,{\rm down})\hs=0.2,\\
  P(x|x,{\rm down})\hs=0.8,\\
  \end{array}
  \right.
 \end{array}
\]
for $x=1$ and
\[
 \begin{array}{l}
  \left\{
  \begin{array}{ll}
  P(x+1|x,{\rm up})\hs=0.8,\\
  P(x|x,{\rm up})\hs=0.1,\\
  P(x-1|x,{\rm up})\hs=0.1,
  \end{array}
  \right.
  \left\{
  \begin{array}{ll}
  P(x+1|x,{\rm stay})\hs=0.1,\\
  P(x|x,{\rm stay})\hs=0.8,\\
  P(x-1|x,{\rm stay})\hs=0.1,
  \end{array}
  \right.\\
  \left\{
  \begin{array}{ll}
  P(x+1|x,{\rm down})\hs=0.1,\\
  P(x|x,{\rm down})\hs=0.1,\\
  P(x-1|x,{\rm down})\hs=0.8,
  \end{array}
  \right.
 \end{array}
\]
for $x=2,\ldots,16$.
The alarm region is given by $\mc{X}_{\rm a}=\{6,\ldots,16\}$.
The reward function is given by $r_t(x,a)=|x|$ for any $a\in\mc{A}$ and $t\in\mc{T}$.

\begin{figure*}[t]
  \centering
  \subfloat[][]{
    \includegraphics[width=.33\linewidth]{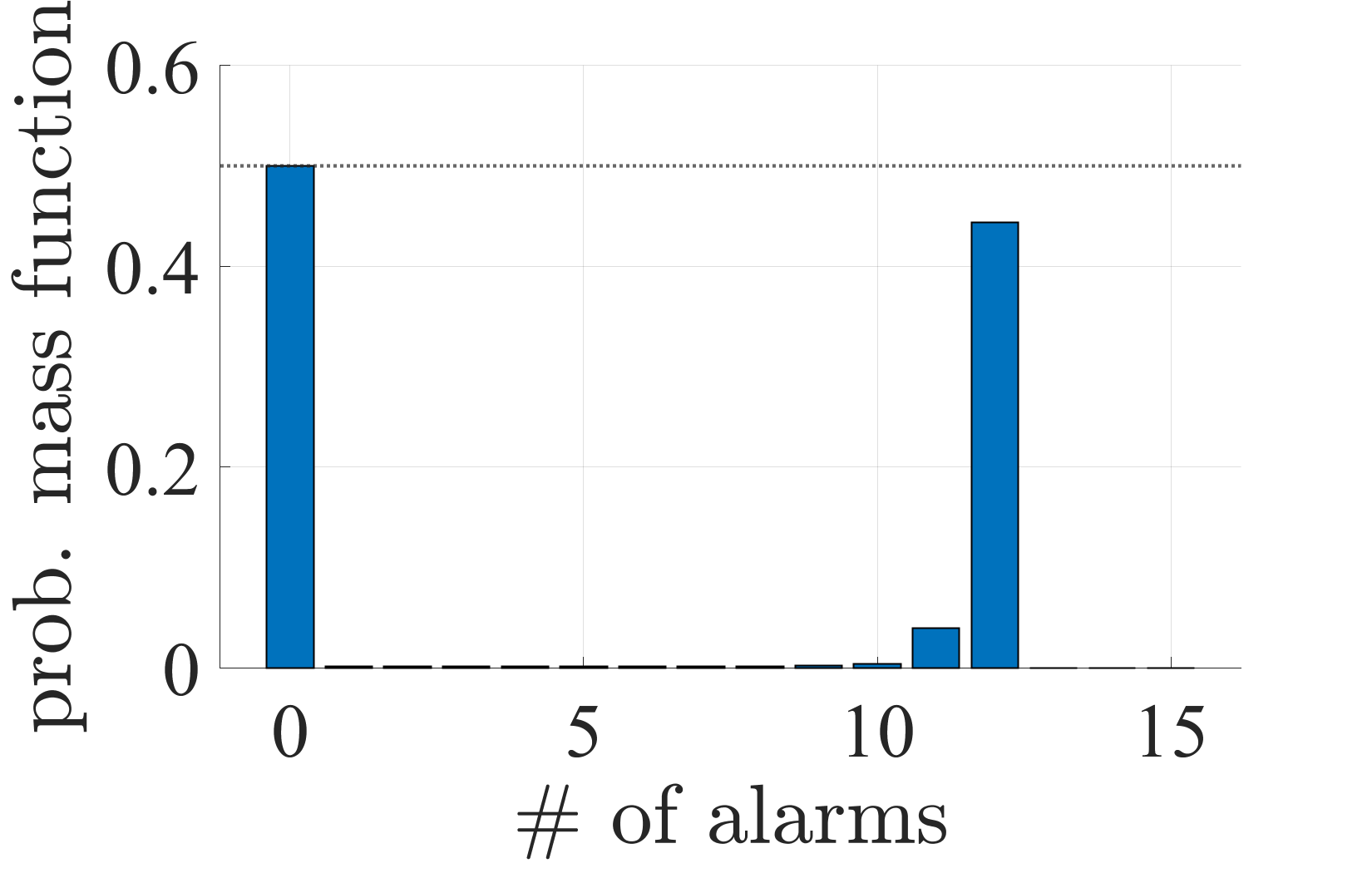}\label{subfig:vio}
    }
  \subfloat[][]{
    \includegraphics[width=.33\linewidth]{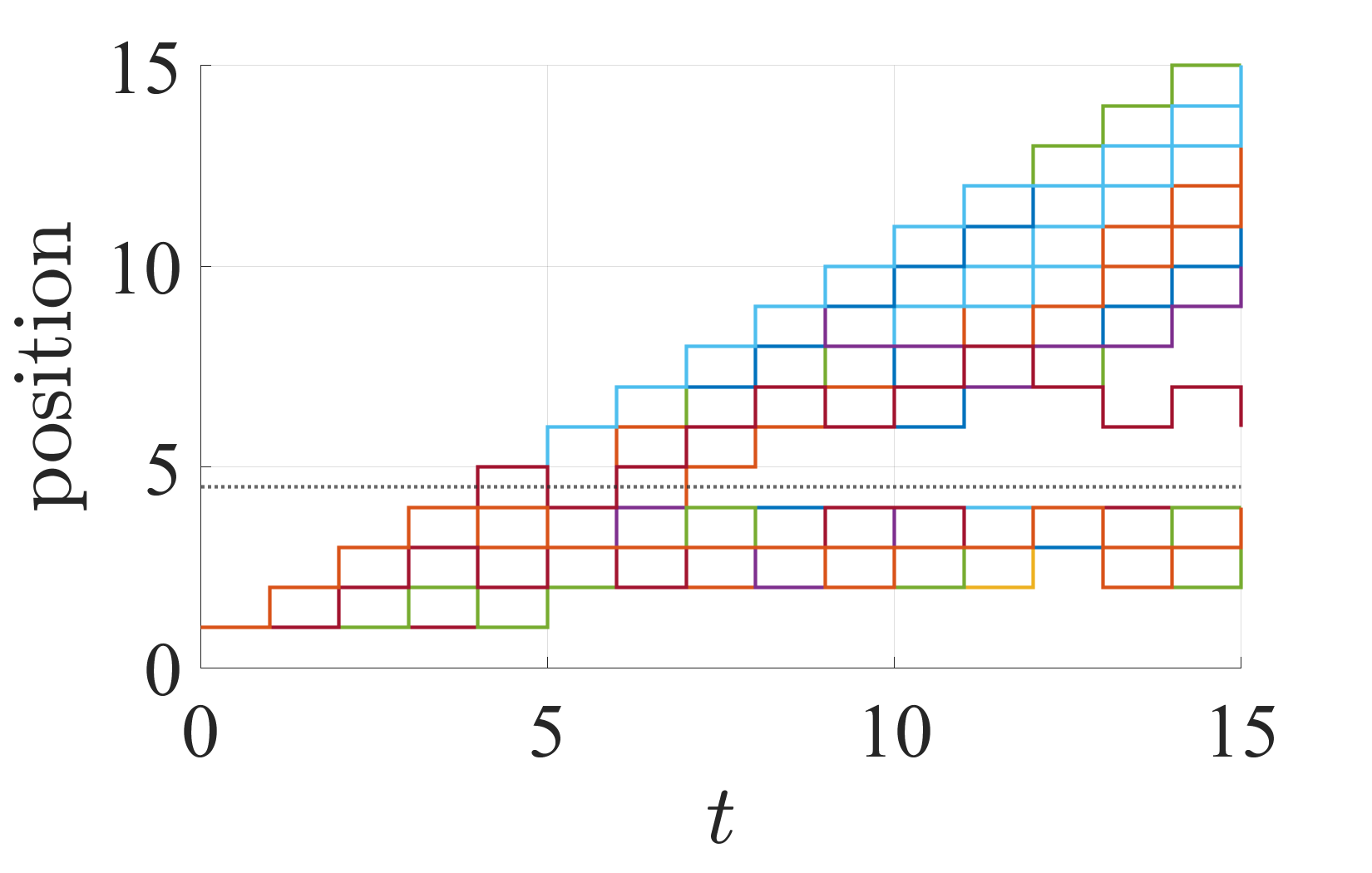}\label{subfig:all}
    }
    \subfloat[][]{
    \includegraphics[width=.33\linewidth]{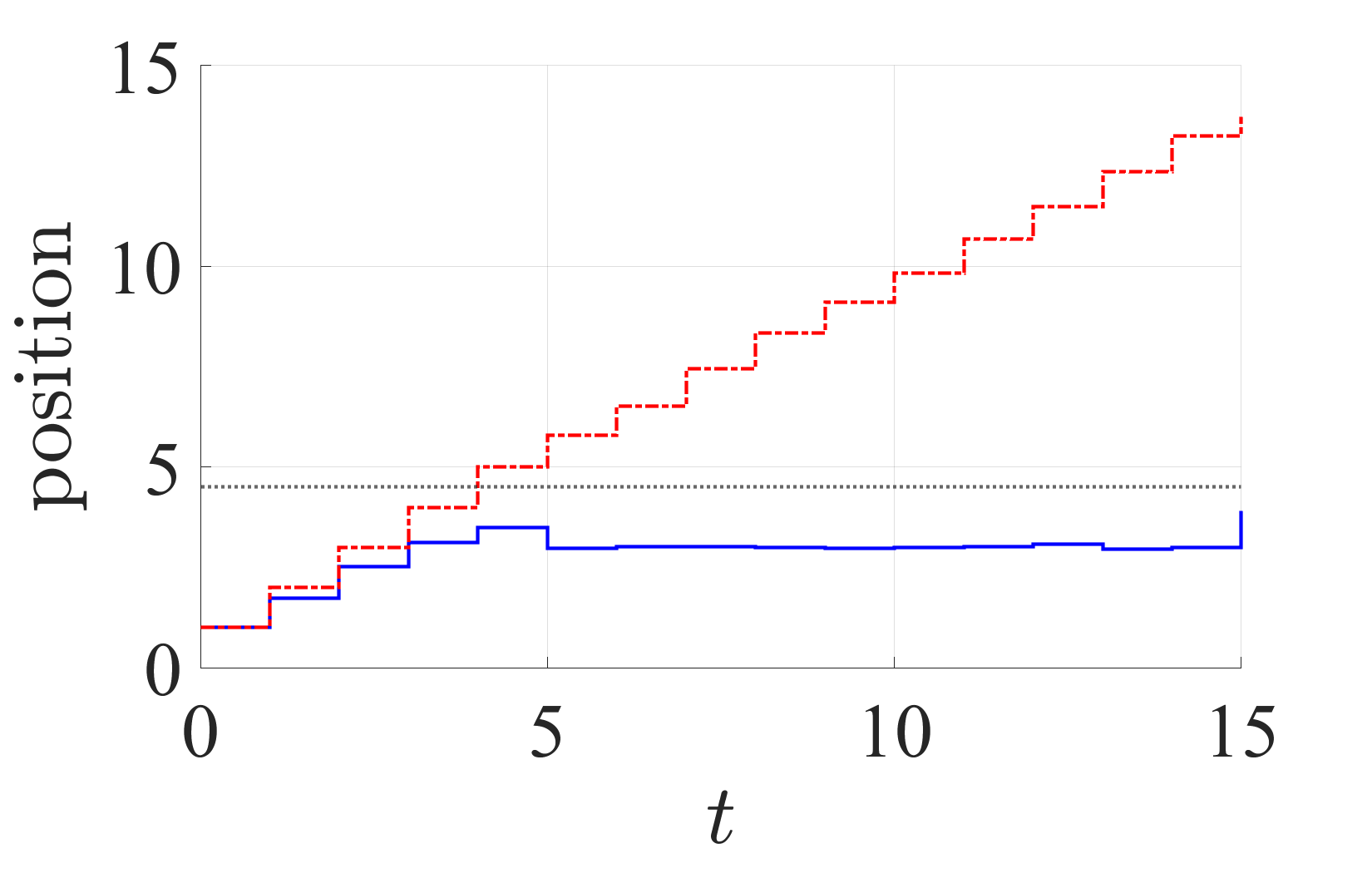}\label{subfig:mean}
    }
    \caption{
    Simulation results with the formulation of Problem~\ref{prob:ori}.
    (a): Probability mass function with respect to the total number of alarms during the process.
    (b): Sample paths with 100 trials. The dotted line describes the boundary of the alarm region.
    (c): Conditional means. The dash line corresponds to the case where the alarm is triggered at least once, while the solid line corresponds to the case where the alarm is not triggered during the process.
    }
    \label{fig:res1}
\end{figure*}

Consider the formulation of Problem~\ref{prob:ori}.
Let the constant on the stealth condition be given by $\Delta=0.5.$
Fig.~\ref{fig:res1} depicts the simulation results with the optimal policy obtained by solving the equivalent formulation~\eqref{eq:LP}.
Fig.~\ref{subfig:vio} depicts the probability mass function with respect to the total number of alarms during the process.
It can be observed that the resulting probability distribution satisfies the stealth constraint in~\eqref{eq:prob_ori} and hence Problem~\ref{prob:ori} is solved through the proposed method.
On the other hand, as pointed out in Sec.~\ref{sec:ex}, the probability is concentrated around the cases more than ten alarms.
This result indicates that the formulation in Problem~\ref{prob:ori} leads to a policy such that the number of alarms becomes large once an alarm is triggered.
Figs.~\ref{subfig:all} and~\ref{subfig:mean} depict the sample paths with 100 trials and the empirical means conditioned by whether an alarm is triggered at least once during the process, or not, respectively.
Those subfigures suggest the same tendency that the adversary does no longer care about being detected once an alarm is triggered.

\begin{figure*}[t]
  \centering
  \subfloat[][]{
    \includegraphics[width=.33\linewidth]{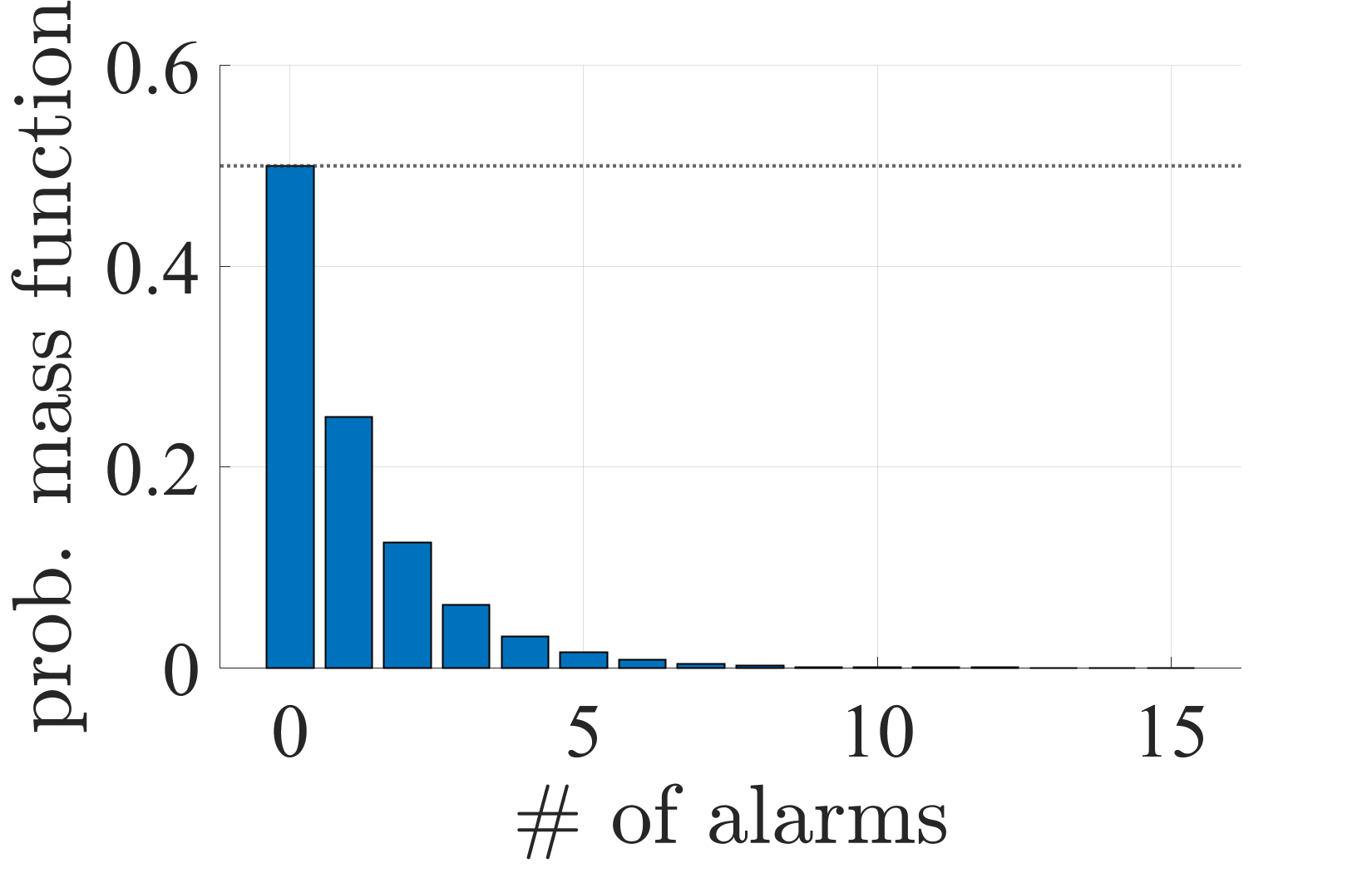}\label{subfig:vio_m}
    }
  \subfloat[][]{
    \includegraphics[width=.33\linewidth]{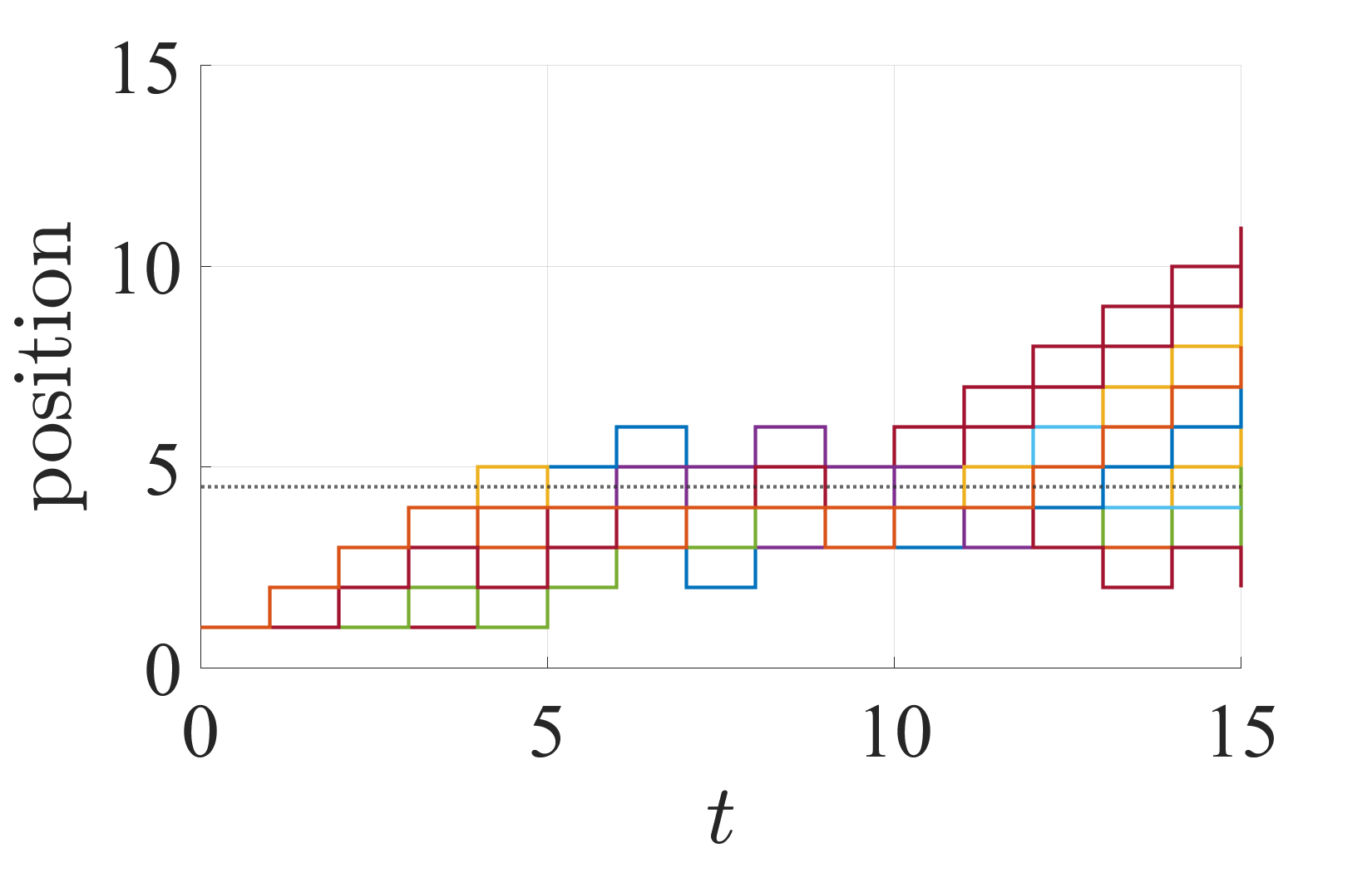}\label{subfig:all_m}
    }
    \subfloat[][]{
    \includegraphics[width=.33\linewidth]{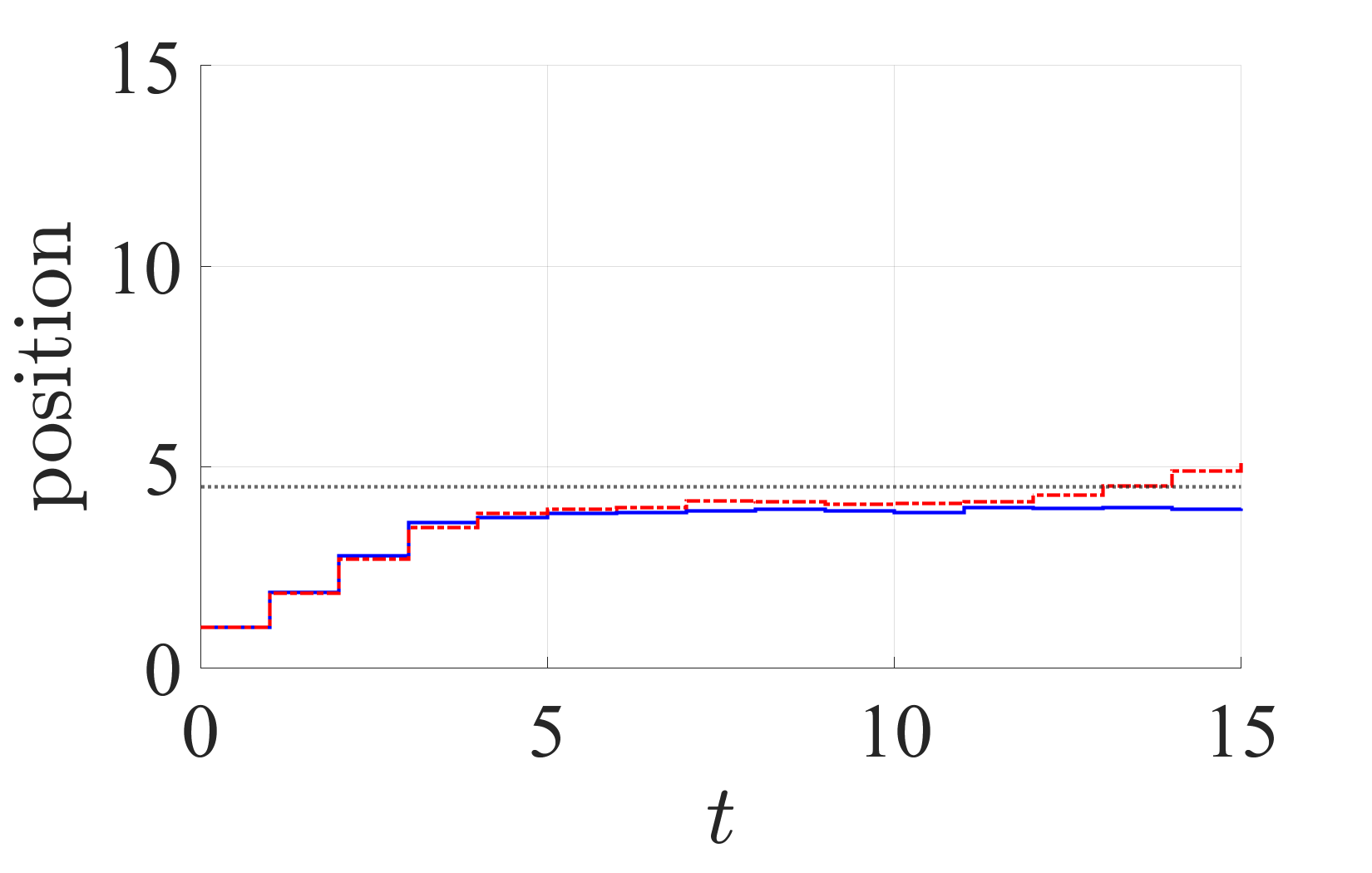}\label{subfig:mean_m}
    }
    \caption{
    Simulation results with the formulation of Problem~\ref{prob:mul}.
    Each subfigure corresponds to those in Fig.~\ref{fig:res1}.
    }
    \label{fig:res2}
\end{figure*}

Next, consider the formulation of Problem~\ref{prob:mul}.
Let the constants on the stealth condition be given by $\Delta_i=(0.5)^i$.
Fig.~\ref{fig:res2} depicts the simulation results.
The subfigures correspond to those in Fig.~\ref{fig:res1}.
It can be observed from Fig.~\ref{subfig:vio_m} that the obtained probability mass function looks more natural than that in Fig.~\ref{subfig:vio}, i.e., the probability of the number of alarms decreases as the number increases.
It can also be observed from Figs.~\ref{subfig:all_m} and~\ref{subfig:mean_m} that the resulting state trajectories are also natural, i.e., those are close to the boundary of the alarm region even if an alarm is triggered during the process.
The evaluated attack impacts through Problems~\ref{prob:ori} and~\ref{prob:mul} are compared in Table~\ref{table:ingredients}.

\begin{table}[t]
\centering
\caption{Evaluated Attack Impacts through Problems~\ref{prob:ori} and~\ref{prob:mul}}
\begin{tabular}{c|c|c}
  & Problem~\ref{prob:ori} & Problem~\ref{prob:mul} \\ \hline
 $J^{\ast}$ & 84.99 & 58.16
\end{tabular}
\label{table:ingredients}
\end{table}

\section{CONCLUSION}
\label{sec:conc}

This study has addressed the attack impact evaluation problem for control systems.
The problem is formulated as an MDP with a temporally joint chance constraint.
The difficulty of the optimal control problem lies in the explosion of the search space owing to the dependency of the optimal policy on the entire history.
In this paper, we have shown that the information whether alarm has been triggered or not is sufficient and the size can be reduced by considering state space augmentation.
Moreover, the problem is extended by taking the number of alarms into account for obtaining reasonable strategies and evaluating the attack impact more precisely.

Further research directions include extension to infinite and continuous state spaces, discretization of which results in a finite but huge state space when the dimension is high.
Moreover, although the size of the search space is reduced compared with the original one, the optimal policy still depends on the time instance.
To reduce its size further and obtain an optimal stationary policy, extension to the infinite horizon problem is of interest.
Finally, in the extended formulation, we have used the full information of the probability distribution.
Clarifying relationships to standard risk measures, such as CVaR, is an open problem.


\appendices
\section{Example: History-Dependency of Optimal Policy and Sufficiency of Alarm Information}\label{app:1}
\begin{figure}[t]
  \centering
  \includegraphics[width=0.98\linewidth]{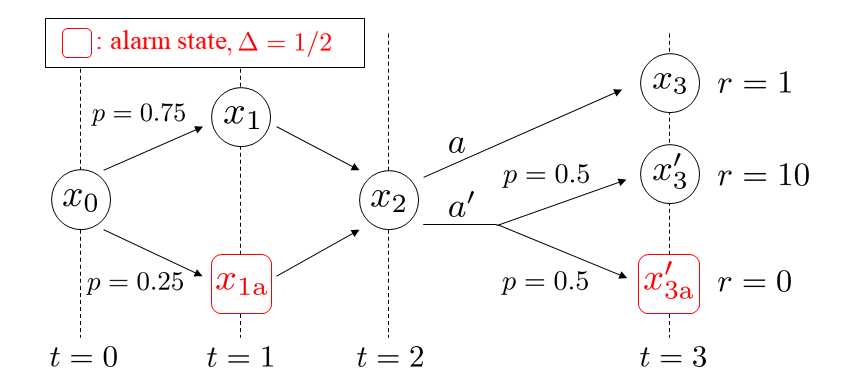}
  \caption{
  MDP for which Markov policies cannot achieve the optimal value of Problem~\ref{prob:ori}.
  The adversary can take an action only at $t=2$ and obtains an reward only at the final step depending only on the state.
  }
  \label{fig:counter_example}
\end{figure}

Consider the MDP illustrated by Fig.~\ref{fig:counter_example}.
In the example, the adversary can take an action only at $t=2$.
When the action $a$ is taken, the state reaches $x_3$ with probability one and the adversary obtains reward $1$.
On the other hand, when the action $a'$ is taken, the state reaches $x'_3$ or $x'_{3{\rm a}}$ with equal probabilities.
The amount of the reward is $10$ in the case of $x'_3$, while no reward is given in the case of $x'_{3{\rm a}}$.
The alarm region is given as $\mc{X}_{\rm a}=\{x_{1{\rm a}},x'_{3{\rm a}}\}$.
The action $a'$ can be interpreted as a risky action in the sense that it leads to large reward in expectation but may trigger an alarm.

The history-dependent policies are parameterized by
$\pih(a'|x_1,x_2)=\alpha$ and $\pih(a'|x_{1{\rm a}},x_2)=\beta$
with parameters $(\alpha,\beta)\in[0,1]\times[0,1]$.
The joint chance constraint is written by
$\Prob^{\pih}(\lor_{t\in\mc{T}}X_t\in\mc{X}_{\rm a})\leq \Delta \Leftrightarrow \Prob^{\pih}(X_1=x_{1{\rm a}}) + \Prob^{\pih}(X_1=x_1)\pih(a'|x_1,x_2)P(x'_{3{\rm a}}|x_2,a')\leq 1/2\Leftrightarrow 1/4+3/4*\alpha/2\leq 1/2\Leftrightarrow \alpha\leq 2/3.$
Thus the feasible region of $(\alpha,\beta)$ is $[0,2/3]\times[0,1]$.
The objective function is written by
$\Prob^{\pih}(X_3=x_3)+\Prob^{\pih}(X_3=x'_3)*10= 3/4*(1-\alpha) + (1-\beta)/4+(3/8*\alpha+\beta/8)*10= 3\alpha+\beta+1.$
Because this is monotonically increasing with respect to $\alpha$ and $\beta$, the optimal values are $(\alpha^{\ast},\beta^{\ast})=(2/3,1),$ which leads to
\[
 \left\{
\begin{array}{ll}
 \pih(a'|x_1,x_2)\hs =2/3,\\
 \pih(a'|x_{1{\rm a}},x_2)\hs =1.
\end{array}
\right.
\]
On the other hand, the Markov policies are parameterized by $\pim(a'|x_2)=\gamma$ with $\gamma\in[0,1]$.
The joint chance constraint is $\gamma\leq 2/3$
and the objective function is $4\gamma+1$.
Thus the optimal value is $\gamma^{\ast}=2/3,$ which leads to $\pim(a'|x_2)=2/3.$
Denoting the value of the objective function with a policy $\pi$ by $J(\pi),$ we have $J(\pih)=4>11/3=J(\pim),$
which implies that Markov policies cannot achieve the optimal value for the example.


The optimal history-dependent policy means that, the adversary reduces the risk when no alarm has been triggered ($X_1=x_1$) while she takes the risky action when there has been an alarm ($X_1=x_{1{\rm a}})$.
In other words, the decision making at the state $x_2$ depends on the alarm history.
This observation leads to the hypothesis that this binary information in addition to the current state is sufficient for optimal decision making and the idea of state space augmentation.

\bibliographystyle{IEEEtran}
\bibliography{sshrrefs}

\end{document}